\documentstyle[epsf,twocolumn,seceq]{orb2001}

\title
{
Mechanisms for Non-Trivial Magnetization Plateaux
of an $\mib{S=}\mbox{\bf 1}$ Frustrated Spin Ladder
}

\def\runtitle
{
Magnetization Plateaux
of an $S=1$ Frustrated Spin Ladder}

\author
{
Kiyomi {\sc Okamoto}, 
Nobuhisa {\sc Okazaki$^{1}$} and
T\^oru {\sc Sakai$^{2}$}
}

\def\runauthor
{
Kiyomi {\sc Okamoto}, 
Nobuhisa {\sc Okazaki} and
T\^oru {\sc Sakai}
}

\inst
{Department of Physics, Tokyo Institute of Technology,
Oh-okayama, Meguro-ku, Tokyo 152-8551, Japan\\
$^1$Kyoto Miyama High School, Murashita, Sasari, Miyama-cho, Kitakuwata-gun,
Kyoto 601-0705, Japan\\
$^2$Tokyo Metropolitan Institute of Technology,
Asahigaoka, Hino, Tokyo 191-0065, Japan
}

\abst
{
We investigate the non-trivial magnetization plateau at $1/4$
of the saturation magnetization of $S=1$ spin ladder,
especially with reference to recent experimental results on
a new organic tetraradical
3,3',5,5'-tetrakis({\it N-tert}-butylaminoxyl)biphenyl, 
abbreviated as BIP-TENO.
We propose three mechanisms for the formation of the plateau;
the N\'eel mechanism, the dimer mechanism and the spin-Peierls
mechanism.
We also discuss the effect of four-spin exchange interactions.
}

\kword
{
magnetization plateau, low-dimensional magnet, spin-Peierls transition
}

\begin{document}
\sloppy
\maketitle


\section{Introduction}

A new organic tetraradical, 
3,3',5,5'-tetrakis({\it N-tert}-butylaminoxyl)biphenyl, 
abbreviated as BIP-TENO, has been synthesized very recently.\cite{Katoh}
This material can be regarded as an $S=1$ antiferromagnetic two-leg spin ladder.
Goto {\it et al.}\cite{Goto} measured the magnetization curve of BIP-TENO 
at low temperatures in high magnetic fields up to about 50 T.
They found $M \simeq 0$ up to about $H_{\rm c1} \simeq 10\,{\rm T}$,
suggesting the existence of the nonmagnetic ground state,
which was consistent with the result of susceptibility measurement.
Another remarkable feature of their magnetization curve is the
existence of the plateau at $M=M_{\rm s}/4$ above $H_{\rm c2} \simeq 45\,{\rm T}$,
where $M_{\rm s}$ is the saturation magnetization.
Unfortunately, the end of the $M_{\rm s}/4$ plateau is unknown
because of the limitation of the strength of their magnetic field.

\begin{figure}[h]
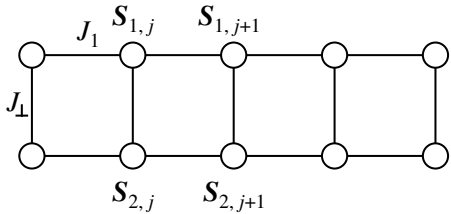

  \vspace{10pt}   
     \epsfxsize=6cm
	    \epsfigure{p66f1.eps} 
  \caption{Simplified model of BIP-TENO.
  Open circles represent $S=1$ spins.}
  \label{fig:S=1}
\end{figure}
A simplified model\cite{Katoh} of BIP-TENO is shown in Fig.~\ref{fig:S=1},
although the real material is more complicated.
Both coupling constants are thought to be antiferromagnetic.
In experimental estimation, it seems $J_\perp \gg J_1$.

Considering the necessary condition for the magnetization
plateau,\cite{OYA}
the spontaneous symmetry breaking along the leg-direction is necessary
for the formation of the plateau.
By use of the model of Fig.~\ref{fig:S=1},
we have already proposed two mechanisms for the plateau;
the N\'eel mechanism and the dimer mechanism.\cite{OOS1,OOS2}

In this paper, we briefly review the N\'eel and dimer mechanism,
and also propose a new possible mechanism --- the spin-Peierls mechanism.
We also discuss the effect of four-spin exchange interactions.

\section{Degenerate Perturbation Theory}

The Hamiltonian of Fig.~\ref{fig:S=1} is expressed as
\begin{eqnarray}
    \hat H
    &=&  J_1 \sum_{l=1,2} \sum_{j=1}^L {\mib S}_{l,j} \cdot {\mib S}_{l,j+1}
     + J_\perp \sum_{j=1}^L {\mib S}_{1,j} \cdot {\mib S}_{2,j} \nonumber \\
    &&~~~~~ - H \sum_{l=1,2} \sum_{j=1}^L S^z_{l,j},
    \label{eq:Ham}
\end{eqnarray}
where ${\mib S}$ denotes the spin-1 operator, 
$j$ the rung number and $l=1,2$ the leg number.
The last term is the Zeeman energy in the magnetic field $H$.
When $J_\perp=0$, our model is reduced to that of
two independent conventional $S=1$ chains.
In this case there appears no plateau in the magnetization curve
except at $M=0$, which corresponds to the Haldane gap.

In the opposite limit $J_\perp \gg J_1$
we can use the degenerate perturbation theory.
When $J_1=0$, all the rung spin pairs are mutually independent.
Thus, at $M=M_{\rm s}/4$, half of the rung spin pairs are
in the state
\begin{equation}
    \psi(0,0) 
    ={1 \over \sqrt{3}}
     \left(\left|\matrix{\uparrow \cr \downarrow} \right\rangle 
      + \left|\matrix{\downarrow \cr \uparrow} \right\rangle
      -\left|\matrix{0 \cr 0}\right\rangle \right),
\end{equation}
and the remaining half pairs are in the state
\begin{equation}
    \psi(1,1)
    = {1 \over \sqrt{2}}
      \left(\left|\matrix{\uparrow \cr 0} \right\rangle
       + \left|\matrix{0 \cr \uparrow} \right\rangle \right),
\end{equation}
where $\psi(S_{\rm tot},S_{\rm tot}^z)$ is the wave function of
with the quantum numbers $S_{\rm tot}$ and $S_{\rm tot}^z$.
These two wave functions have the lowest energies in the subspace of
$S_{\rm tot}^z = 0$ and $S_{\rm tot}^z = 1$, respectively.
The $M_{\rm s}/4$ state is highly degenerate as far as $J_1=0$,
because there is no restriction for the configurations of these two states. 
This degeneracy is lifted up by the introduction of $J_1$.
To investigate the effect of $J_1$, we introduce the pseudospin ${\mib T}$ with
$T=1/2$.
The $|\Uparrow\rangle$ and $|\Downarrow\rangle$ states of the ${\mib T}$ spin
correspond to $\psi(1,1)$ and $\psi(0,0)$, respectively.
Neglecting the other seven states for the rung spin pairs,
the effective Hamiltonian can be written as, 
through straightforward calculations
\begin{equation}
    \hat H_{\rm eff}
    = \sum_j \left\{  J_{\rm eff}^{xy} (T_j^x T_{j+1}^x + T_j^y T_{j+1}^y)
                    + J_{\rm eff}^z T_j^z T_{j+1}^z \right\},
    \label{eq:T-Ham}
\end{equation}
in the lowest order of $J_1$, where
\begin{equation}
    J_{\rm eff}^{xy} = (8/3)J_1,~~~~~
    J_{\rm eff}^z = (1/2)J_1,
    \label{eq:Del}
\end{equation}
Thus, the $M_{\rm s}/4$ plateau problem of the original model
in a magnetic field is mapped onto the $M=0$ problem
of the $T=1/2$ antiferromagnetic
$XXZ$ spin chain with the nearest-neighbor (NN) interaction
in the absence of magnetic field.
As is well known, 
depending on whether 
$\Delta_{\rm eff} \equiv J_{\rm eff}^z/J_{\rm eff}^{xy} \le 1$
or $\Delta_{\rm eff} >1$, 
the ground state of eq. (\ref{eq:T-Ham}) is either the
spin-fluid state (gapless) or the N\'eel state (gapful),
which correspond to the no-plateau state or the plateau state of the
original model at $M_{\rm s}/4$, respectively.
Since $\Delta_{\rm eff} = 3/16$ in the present case,
no $M_{\rm s}/4$ plateau is expected as far as eq.(\ref{eq:T-Ham})
can be applied.
As already stated, neither is there an $M_{\rm s}/4$ plateau when $J_\perp=0$.
Thus, the $M_{\rm s}/4$ plateau will not appear in the entire region of
$J_1/J_\perp$ for the model of Fig.~\ref{fig:S=1}.
We have checked this conclusion by numerical methods.

\begin{figure}[h]
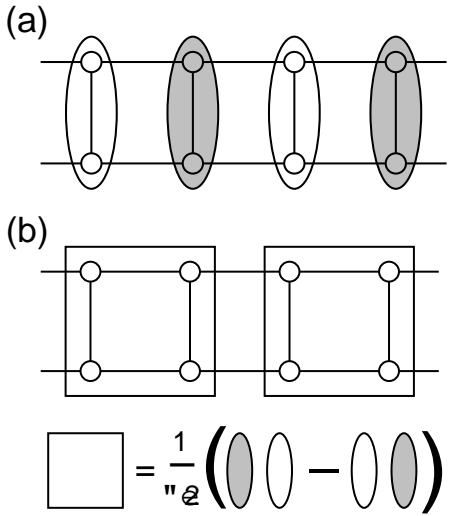

  \vspace{10pt}   
     \epsfxsize=6cm
     \epsfigure{p66f2.eps} 
     \epsfxsize=6cm 
     \epsfigure{p66f3.eps}
  \caption{(a) Physical picture of N\'eel mechanism plateau.
Open and shaded ellipses represent $\psi(0,0)$ and $\psi(1,1)$
states of the rung pairs ($|\Downarrow\rangle$ and $|\Uparrow\rangle$ in the
$\mib T$-picture), respectively.
(b) Physical picture of dimer mechanism plateau.}
  \label{fig:1}
\end{figure}

If we add frustrated spin interactions 
\begin{eqnarray}
    &&J_2\sum_i^L({\mib S}_{1,i}
    \cdot {\mib S}_{2,i+1}+{\mib S}_{2,i} \cdot {\mib S}_{1,i+1}) \nonumber \\
    &&+J_3\sum_i^L({\mib S}_{1,i}
    \cdot {\mib S}_{1,i+2}+{\mib S}_{2,i} \cdot {\mib S}_{2,i+2}),
\end{eqnarray}
to the Hamiltonian (\ref{eq:Ham}),
the coupling constants in the effective Hamiltonian
are modified to
\begin{equation}
    J_{\rm eff}^{xy} = (8/3)(J_1-J_2),~~~~
    J_{\rm eff}^z = (1/2)(J_1+J_2),
\end{equation}
and also there appear the next-nearest-neighbor interactions
\begin{equation}
\frac{8J_3}{3}\sum_i^L({T}^x_{i} \cdot {T}^x_{i+2}
+{T}^y_{i} \cdot {T}^y_{i+2})  
+\frac{J_3}{2}\sum_i^L {T}^z_{i} \cdot {T}^z_{i+2}.
\label{eq:T-ham2}
\end{equation}
Thus our effective Hamiltonian is of the generalized version of
the $T=1/2$ $XXZ$ chain with the next-nearest-neighbor interactions.
In this case there is a possibility of the N\'eel state and the dimer state 
of the $\mib T$-system.\cite{NO2}
Although both states are gapped states
(plateau states in the $\mib S$-picture),
the physical picture of the plateau formation is quite different
with each other.

We have already shown\cite{OOS1} that the condition for the realization of
the N\'eel plateau is $J_2/J_1 \ge 13/19$ when $J_3=0$.
Also the dimer plateau condition is $J_3/J_1 \ge 0.31$
when $J_2=0$.

\section{Spin-Peierls Mechanism}

Another possible mechanism is the spin-Peierls mechanism.
As is well known, 
usual spin-Peierls transition is observed in $\tilde S=1/2$
isotropic Heisenberg chain at $M=0$.
The uniform nearest-neighbor spin interaction
$J {\tilde {\mib S}}_j \cdot {\tilde {\mib S}}_{j+1}$
is modified into the alternating form 
$J (1+\delta)^j {\tilde {\mib S}}_j \cdot {\tilde {\mib S}}_{j+1}$
due to the lattice distortion, 
where $\delta$ measures the magnitude of the bond-alternation.
Since the dimer correlation behaves as $r^{-1}$ in the long distance limit
for the uniform case (quasi long-range order),
the uniform chain is very sensitive to the bond alternation.
In fact, the energy gain of the spin system due to the bond-alternation
behaves as\cite{CF} $\delta^a$ with $a=4/3$, 
where $a$ is the energy gain exponent..
Here we have omitted the logarithmic correction.\cite{BE}  
Because the energy loss of the lattice system is proportional to $\delta^2$,
there occurs the bond-alternation induced by spontaneous
lattice distortion.
Two spins on a stronger bond form a singlet pair effectively.
Thus there is a finite energy gap in the excitation spectrum,
which corresponds to the formation energy of an effective singlet pair.

\begin{figure}[h]
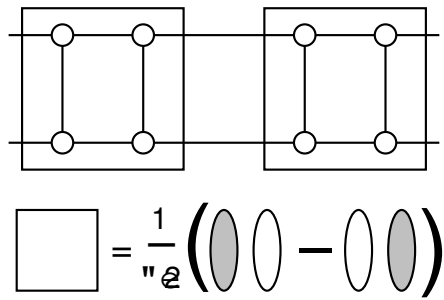

  \vspace{10pt}   
     \epsfxsize=6cm
	    \epsfigure{p66f4.eps} 
  \caption{Physical picture of the spin-Peierls mechanism.
           The distance between neighboring two spins on a leg
           changes alternatingly.
           This is the origin of the bond-alternation.}
  \label{fig:sp}
\end{figure}

Since the original Hamiltonian can be mapped onto the $T=1/2$ $XXZ$ chain,
a similar spin-Peierls scenario is also possible for the present model.
The physical picture of the spin-Peierls mechanism is shown in
Fig.~\ref{fig:sp}.
The energy gap in the excitation spectrum of the ${\mib T}$ system
corresponds to the magnetization plateau at $M_{\rm s}/4$
of the original system.
Of course, the value of the energy gain exponent $a$
should be modified from $4/3$,
because our effective Hamiltonian is no longer isotropic 
(now $\Delta_{\rm eff} = 3/16 \ne 1$).
The $\Delta_{\rm eff}$ dependence of $a$ is\cite{NF,NO}
\begin{equation}
    a = {4 \over 4-\eta},~~~~~
    \eta = {2 \over 1 + (2/\pi) \cos^{-1} \Delta_{\rm eff}}
\end{equation}
which results in
\begin{equation}
    a = 1.80 < 2
    \label{eq:a}
\end{equation}
for our case.
Thus the spin-Peierls mechanism is possible from
the consideration of the energy.
Our preliminary numerical calculation based on the $S=1$ model
shows that the $a$ is very close to $1.8$
as conjectured in eq. (\ref{eq:a}).

\section{Effect of Four-Spin Cyclic Exchange Interactions}

\begin{figure}[h]
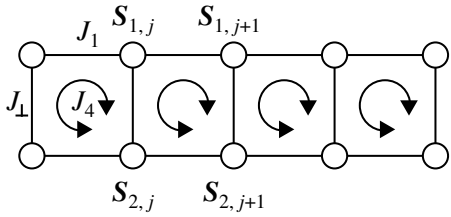

  \vspace{10pt}   
     \epsfxsize=6cm
	    \epsfigure{p66f5.eps} 
  \caption{Four-spin cyclic exchange interactions.}
  \label{fig:4-spin}
\end{figure}

Let us consider the effect of four-spin cyclic exchange interactions
shown in Fig. \ref{fig:4-spin}.
In the $S=1/2$ two-leg ladder, this interaction can
brings about the non-trivial magnetization plateau.\cite{Nakasu}
The four-spin cyclic exchange interaction works as
\begin{eqnarray}
    &&P\left( \matrix{\alpha &\beta \cr \gamma &\delta}\right)
    =  \left( \matrix{\beta  &\delta \cr \alpha &\gamma}\right), \\
    &&P^{-1}\left( \matrix{\alpha &\beta \cr \gamma &\delta}\right)
    =  \left( \matrix{\gamma &\alpha \cr \delta &\beta}\right).
\end{eqnarray}
where $\alpha,\beta,\gamma,\delta$ are the related four spins.
If we add this interaction with magnitude $J_4$
to the Hamiltonian (\ref{eq:Ham}),
the effective coupling constants in eq. (\ref{eq:Del}) are modified to
\begin{equation}
    J_{\rm eff}^{xy} = {8J_1 + 4J_4 \over 3},~~~
    J_{\rm eff}^z    = {3J_1 + 2J_4 \over 6}.
\end{equation}

Here we do not take the $J_2$ and $J_3$
interactions into consideration
for simplicity.
\vspace{10cm}
It is easy to see that the N\'eel state in the $\mib T$-system
is not realized for any choice of $J_4$.
Thus the four-spin cyclic exchange cannot bring about the
plateau in our problem.
This situation is quite different from that in the $S=1/2$
two-leg ladder.\cite{Nakasu}

\section{Summary}

We have proposed 
three mechanisms for the plateau of $S=1$ ladder,
BIP-TENO.
These mechanisms are directly distinguished by experimental methods
in principle, although they might be difficult to realize.
In the N\'eel plateau state,
the expectation values of $S^z$ of spins along a leg are
$1/2,0,1/2,0,1/2,\cdots$,
although this \lq\lq complete N\'eel configuration\rq\rq is
somewhat smeared by the quantum fluctuations.
On the other hand, in the dimer plateau state,
they are $1/4,1/4,1/4,1/4,1/4,\cdots$.
In the spin-Peierls plateau,
the spontaneous lattice distortion can be observed,
although the spin configurations is the same
as those in the dimer plateau state.

Because such experiments seem to be difficult in high magnetic fields,
we have to surmise the nature of the plateau state from the
behaviors of the magnetization curve and the magnetic susceptibility,
by comparing them with the analytical and numerical results.
The investigation in this line is now in progress.\cite{OOS2}





\end{document}